\documentstyle[12pt,epsf]{article}

\newcommand{\be}{\begin{equation}}
\newcommand{\ee}{\end{equation}}
\newcommand{\eel}[1]{\label{#1}\end{equation}}
\newcommand{\bea}{\begin{eqnarray}}
\newcommand{\eea}{\end{eqnarray}}
\newcommand{\eeal}[1]{\label{#1}\end{eqnarray}}
\newcommand{\baq}{\begin{equation}\begin{array}{rcl}}
\newcommand{\eaq}{\end{array}\end{equation}}
\newcommand{\eaql}[1]{\end{array}\label{#1}\end{equation}}
\newcommand{\beac}{\begin{equation}\begin{array}{rcl}}
\newcommand{\eeacn}[1]{\end{array}\label{#1}\end{equation}}
\newcommand{\ba}{\begin{array}}
\newcommand{\ea}{\end{array}}
\newcommand{\non}{\nonumber \\}

\newcommand{\al}{{\alpha^{'}}}
\newcommand{\beq}{\begin{eqnarray}}
\newcommand{\eeq}{\end{eqnarray}}

\newcommand{\gym}{g_{YM}}

\begin{document}
\newcommand{\preprint}[1]{\begin{table}[t]  
           \begin{flushright}               
           \begin{large}{#1}\end{large}     
           \end{flushright}                 
           \end{table}}                     

\begin{flushright}
hep-th/9903067\\
NSF-ITP-99-012
\end{flushright}

\begin{center}
\Large{{\bf A Comment on the Zamolodchikov c-Function\\ and the Black
String Entropy}}

\vspace{10mm}

\normalsize{Akikazu Hashimoto}

\vspace{10mm}

Institute for Theoretical Physics\\ University of California,
Santa Barbara, CA 93106\\
aki@itp.ucsb.edu\\

\vspace{10mm}

N. Itzhaki\\

\vspace{10mm}

Department of Physics\\ University of California,
Santa Barbara, CA 93106\\
sunny@physics.ucsb.edu\\

\end{center}

\vspace{10mm}

\begin{abstract}
\noindent Using the spectral representation approach to the
Zamolodchikov's c-function and the Maldacena conjecture for the
D1-branes, we compute the entropy of type IIB strings.  An agreement,
up to a numerical constant which cannot be determined using this
approach, with the Bekenstein-Hawking entropy is found.
\end{abstract}

\newpage

\baselineskip 18pt

In two dimensional field theories, the two-point function of the
energy-momentum tensor is a useful concept in studying the relation
between the energy and the entropy.  For conformal theories, they
completely determine the entropy \cite{car}.  This fact, together with
the observation of Brown and Henneaux \cite{bh} that the asymptotic
group of $AdS_3$ yields a two dimensional conformal theory, was used
by Strominger to derive the Bekenstein-Hawking entropy for black holes
in $AdS_3$ \cite{str}.

In this note we study the SYM theory in 1+1 dimensions with gauge
group $SU(N)$ and sixteen supercharges which is a non-conformal
theory. This theory can be thought of as the theory living on a
collection of $N$ D1-branes in the low energy ``decoupling'' limit. In
the extreme UV the theory is free and conformal with central charge
\be\label{a}
c_{UV}\sim N^2.
\ee
Perturbation theory in SYM can be trusted in the UV as long as the
effective coupling constant is small, that is
\be
1\gg g_{eff}^2=\gym^2 N x^2 ~~\Longrightarrow ~~ x\ll \frac{1}{\gym \sqrt{N}},
\ee
where $x$ is the scale being probed by the two point function.  In the
deep IR region, the physical energy scale determined by the coupling
constant becomes irrelevant and the theory flows to a conformal
theory.  In \cite{hms,bjsv} this theory was shown to be a conformal
$\sigma$-model with the target space $(R^8)^N/S_N$ whose central
charge is
\be\label{b}
c_{IR}\sim N.
\ee
Note that $c_{UV}>c_{IR}$ as expected from the Zamolodchikov's
c-theorem \cite{zam}.  The first correction to the orbifold CFT is
given by the twist operator $ \frac{1}{\gym} V_{ij}$ where $i, j$
label the fields on which the twist operator is acting \cite{dvv}.
There are various ways to show that the perturbation theory with
respect to the twist operators breaks down at
$x<\frac{\sqrt{N}}{\gym}$ \cite{dvv2,juan,pp}.  A simple argument
which rests on the c-theorem is the following.  Perturbation theory
around the conformal point will break down when the difference between
the Zamolodchikov c-function and $c_{IR}$ is of the order of $c_{IR}$.
That is, when $\langle T_{\bar{z}z}(x) T_{\bar{z}z}(0) \rangle$ is of
the order of $\langle T_{zz}(x) T_{zz}(0) \rangle$.  Conservation of
the energy-momentum tensor implies that (see e.g.~\cite{joe})
\be\label{b1}
T_{\bar{z}z}= -\frac{\pi \sum V_{ij}}{\gym}.
\ee
Therefore, perturbation theory around the conformal 
point can be trusted when
\be
\frac{c_{IR}}{x^4}\gg \frac{N^2}{\gym^2 x^6},~~\Longrightarrow ~~ x
\gg\frac{\sqrt{N}}{\gym},
\ee
where we have used the fact that the weight of the twist operators
 is $(3/2, 3/2)$ \cite{dvv}.

In the large $N$ limit there is a large region, 
\be\label{66}
\frac{1}{\sqrt{N}\gym}\ll x\ll \frac{\sqrt{N}}{\gym}, \ee
for which neither the perturbative SYM nor the orbifold CFT
description can be trusted.  In \cite{juan} it was shown that this
region is best described by the type IIB string theory on a background
associated with the near horizon geometry of D1/F1 strings.  Exactly
at the points in UV and the IR regions where the perturbative field
theory descriptions break down, the curvature (in string units) is
 small so that the supergravity approximation
becomes reliable.  The transitions between the perturbative conformal
theories (at the UV and IR) and the supergravity description were
studied in \cite{gj,juan1,andi} to find a match from both sides up
to a numerical coefficient which cannot be determined using current
methods.  This agreement (for the entropy \cite{gj,andi} and for the
Wilson line \cite{juan1,andi}) supports the Maldacena conjecture for
this non-conformal theory but it does not give us much information
about the way in which the supergravity description interpolates
between the UV and the IR perturbative field theories descriptions.

In this article, we elaborate on the interpolation of the
Zamolodchikov's c-function between $c_{IR}$ and $c_{UV}$ and its
relation to the entropy of the near-extremal D1/F1 string. To make
contact with entropy it is useful to use the Kallen-Lehmann spectral
representation of the correlator of two energy-momentum tensors
\cite{cfl},
\be\label{1}
\langle T_{\mu\nu}(x)T_{\rho\sigma}(0)\rangle=\frac{\pi}{3}\int_0^{\infty}d\mu
 c(\mu)
\int\frac{d^2p}{(2\pi)^2}e^{ipx}\frac{(g_{\mu\nu}p^2-p_{\mu}p_{\nu})
(g_{\rho\sigma}p^2-p_{\rho}p_{\sigma})}{p^2+\mu^2}.
\ee
The fact that the Zamolodchikov's c-function is monotonically
decreasing along the RG flow follows from $c(\mu)\geq 0$ which must
hold for any unitary theory \cite{cfl}.  In two dimensions, covariant
quantity with four indices subject to the constraint following from
the conservation of energy momentum-tensor is characterized by a
single invariant. Thus, there is only one possible function of the
intermediate mass scale, $c(\mu)$, which is known as the spectral
density. The quantity
\be c_{eff}(\Lambda) = \int_0^\Lambda d\mu\,  c(\mu), \ee
interpolates between $c_{IR} = c_{eff}(0)$ and $c_{UV} =
c_{eff}(\infty)$.  Since $c(\mu)d\mu$ measures the density of degrees
of freedom which couple to the energy-momentum tensor, and since all
fields couple to the energy-momentum tensor, the spectral
representation of the correlator of two energy-momentum tensors
measures the density of degrees of freedom.

Using the methods developed in \cite{gkp,witten} to compute field
theory correlation functions via the bulk propagation of supergravity
modes, we can calculate the two point function of the energy-momentum
tensor.  Suppressing numerical factors and Lorentz indices, we find
\be\label{2} \langle T(x)T(0) \rangle=\frac{N^{3/2}}{\gym x^5}.  \ee
Before substituting this into eq.(\ref{1}) and discussing the spectral
density, it is worth while to make a few comment about this result.
Eq.(\ref{2}) is obtained by repeating the procedure of
\cite{gkp,witten} for the minimally coupled scalar in the near horizon
geometry of the D1-brane.  In  non-conformal theories
 it is harder to identify the correspondence between the   
supergravity modes and the field theory operators since the symmetry
group is smaller (see however \cite{jky}).  General covariance
indicates that the energy momentum tensor must correspond to the
metric fluctuation $h_{\mu\nu}$.  It is therefore more appropriate to
analyze the field equations for metric fluctuations in this
background. We expect nonetheless for the generic components of the
metric fluctuations to behave essentially like a minimal scalar.  The
reason is that $h_{\mu}^{\mu}$ mixes with a linear combination of a
minimal and fixed scalar\footnote{Strictly speaking, these fields are
scalars in the 9 dimensional supergravity obtained by dimensionally
reducing along the spatial direction of the D1-brane.}
\cite{fixed}. In the supergravity region the fixed scalar contribution
is suppressed and we are left with eq.(\ref{2}) for $\langle
T_{\bar{z}z}(x) T_{\bar{z}z}(0) \rangle$. All other components are
determined in two-dimensions by the conservation of the
energy-momentum tensor.

There are corrections to (\ref{2}) suppressed by $\frac{1}{\gym
x\sqrt{N}}$ and $\frac{x \gym}{\sqrt{N}}$ which can be thought of
respectively as curvature and quantum corrections from the point of
view of type IIB string theory in the near horizon geometry of the
D1-brane. These corrections are very small and can be ignored in the
region given by eq.(\ref{66}) where supergravity approximation can be
trusted. Conversely, the point in $x$-space where these corrections
become significant mark the transition point to the UV and IR
conformal fixed points.  At these transition points, between
supergravity and perturbative SYM and between supergravity and the
orbifold CFT, eq.(\ref{2}) agrees (up to a numerical factor) with the
conformal results
\be
\langle T(x)T(0) \rangle =\frac{c}{x^4},
\ee
for the central charge appropriate for the UV and IR fixed points
given in eqs.(\ref{a},\ref{b}). In order to fix the numerical
coefficient of eq.(\ref{2}) unambiguously, it may be  necessary to
understand the supergravity-perturbative SYM crossover in detail so
that the normalization in the supergravity region can be matched to
the normalization in the perturbative SYM region. For our purpose,
however, there is no need to fix the numerical constant since the
relation between the spectral density and the entropy can be
determined only up to a numerical factor as we discuss below.

Combining eq.(\ref{2}) with eq.(\ref{1}) we find that $c(\mu)=N^{3/2}/\gym$.
  Therefore, for a given temperature $\hat{T}$,
the number of light degrees of freedom with $p^2<\hat{T}^2$ is
\be {\cal N}_{eff}(\hat{T}) \sim c_{eff}(\hat{T}) = \frac{N^{3/2}
\hat{T}}{\gym}. \ee
%
Here we encounter a fundamental ambiguity: the relative numerical
coefficient between $c_{eff}(\hat T)$ and ${\cal N}_{eff}(\hat T)$
cannot be determined for non-conformal theories\footnote{For an
attempt to go off criticality see \cite{cond}.} (as opposed to
conformal theories where ${\cal N}_{eff}=c_{eff}$).  In fact, one can
construct two theories with the same $c_{eff}$ whose ${\cal N}_{eff}$
agrees only up to a numerical factor of order one.

The contribution to the free energy is 
\be
F \sim {\cal N}_{eff}(\hat{T}) L \hat{T}^2.
\ee
Hence the energy density and entropy density are
\beq\label{7}
&& s=\frac{S}{L}=\frac{N^{3/2} \hat{T}^2}{\gym},\non &&
\epsilon=\frac{E}{L}=\frac{N^{3/2} \hat{T}^3}{\gym}.
\eeq

We wish to compare this with the black hole thermodynamics.  In the
Einstein frame the near horizon metric of $N$ near-extremal D1-branes
is,
\be
\frac{ds^2}{\al}=\frac{U^{9/2}}{\gym^{5/2} N^{3/4}}\left(
-\left (1-\frac{U_0^6}{U^6} \right)dt^2
+dx^2\right) +
\frac{ N^{1/4}}{\sqrt{\gym}U^{3/2}(1-\frac{U_0^6}{U^6})}dU^2
+\frac{N^{1/4}\sqrt{U}}{\sqrt{\gym}}d\Omega_6^2,
\ee
where $U_0^6=\gym^4 \epsilon$.  This yields for the entropy density
\cite{juan,kt},
\be\label{ab}
s=\gym^{-1/3}\sqrt{N}\epsilon^{2/3},
\ee
which is in agreement, up to a numerical factor, with eq.(\ref{7}).

Note that unlike in the near horizon geometry of D5+D1 branes, the
field theory describing string theory in the near horizon geometry of
D1-branes is known in details: it is the SYM in two dimensions.
However, our calculation did not rely on the detailed properties of the
SYM action at all. What we did instead was to use the supergravity
dual to compute a field
theory quantity, the two-point function of the energy-momentum
tensor. Then, using rather general field theory arguments which are
valid for any unitary field theory in two dimensions, we compute the
entropy to find an agreement with the Bekenstein-Hawking formula. This
agreement serves as a check of Maldacena conjecture for D1-branes
\cite{juan}.  

It is interesting to note that in the supergravity region the entropy
energy relation is similar to that of a gas of a free massless scalar
field in (2+1) dimensions. In the extreme UV and IR, on the other
hand, this theory behaves like a free gas in (1+1) dimensions (with a
different number of field, since $c_{UV} > c_{IR}$).  Amusingly, this
general behavior is mimicked by the following simple statistical
mechanical model.  Consider $N$ free fields in (2+1) dimensions which
propagate on a ``semi-lattice.''  By ``semi-lattice'' we mean a chain
of $\hat{N}$ continuous strings with lattice spacing $a$.  So the size
of the system is $L_1 L_2$ where $L_1$ is an IR cutoff which we can
take to infinity and $L_2=\hat{N}a$.  The dispersion relation for this
system is 
\be
\label{iop} \omega ^2=k_1^2+ \frac{4}{a^2} \sin ^2 (k_2
a/2), ~~~~k_2=n\pi/L_2,~~~~ n=0...\hat{N}.  
\ee 
In the IR where the
thermal wave-length is larger than $L_2$, the contributions from the
extra dimension are negligible and we have the entropy of a free
massless gas in two dimensions.  In the UV where the thermal
wave-length is smaller than $a$, $\omega$ can grow only due to $k_1$
as $k_2$ is restricted to a single Brillouin zone.  Therefore, the
entropy is that of a free massless gas in two dimensions where the
number of massless fields is $N N^{'}$ as can be read from
eq.(\ref{iop}).\footnote{The fields are not exactly massless as can be
seen from eq.(\ref{iop}). Rather their masses are bounded by $1/a$
which is much smaller then the temperature so their contribution to
the entropy at large temperature is similar to the massless fields
contributions.}  To have an agreement with eq.(\ref{a}) we set
$\hat{N}=N$.  In the intermediate region the system behaves like a gas
of $N$ free particles in (2+1) dimensions. To match with the number of
degrees of freedom of 1+1 dimensional SYM in the deep UV and IR, we
set $a=\frac{\sqrt{N}}{\gym}$.

\section*{Acknowledgments}

We would like to thank Ilya Gruzberg, Victor Gurarie and Joe
Polchinski for helpful discussions. AH is supported in part by the
National Science Foundation under Grant No. PHY94-07194. NI is
supported in part by the NSF grant PHY97-22022.

\end{document}